\documentclass[conference]{IEEEtran}
\IEEEoverridecommandlockouts
\usepackage{cite}
\usepackage{amsmath,amssymb,amsfonts}
\usepackage{algorithmic}
\usepackage{xwatermark}
\usepackage{graphicx}
\usepackage{textcomp}
\usepackage{lipsum}
\usepackage{xcolor}
\def\BibTeX{{\rm B\kern-.05em{\sc i\kern-.025em b}\kern-.08em
    T\kern-.1667em\lower.7ex\hbox{E}\kern-.125emX}}
\begin{document}
\IEEEoverridecommandlockouts
\IEEEpubid{\makebox[\columnwidth]{978-1-6654-8939-3/22/\$31.00~\copyright2022 IEEE\hfill} \hspace{\columnsep}\makebox[\columnwidth]{ }}
\title{Sequential Embedding-based Attentive (SEA) classifier for malware classification\\
}
\IEEEpubidadjcol

\author{\IEEEauthorblockN{1\textsuperscript{st} Muhammad Ahmed}
\IEEEauthorblockA{\textit{Department of Computer Science} \\
\textit{National University of Computer and Emerging Sciences}\\
Karachi, Pakistan \\
k180231@nu.edu.pk}
\and
\IEEEauthorblockN{2\textsuperscript{nd} Anam Qureshi}
\IEEEauthorblockA{\textit{Department of Computer Science} \\
\textit{National University of Computer and Emerging Sciences}\\
Karachi, Pakistan \\
anam.qureshi@nu.edu.pk}
\and
\IEEEauthorblockN{3\textsuperscript{rd} Jawwad Ahmed Shamsi}
\IEEEauthorblockA{\textit{Department of Computer Science} \\
\textit{National University of Computer and Emerging Sciences}\\
Karachi, Pakistan \\
jawwad.shamsi@nu.edu.pk}
\and
\IEEEauthorblockN{4\textsuperscript{th} Murk Marvi}
\IEEEauthorblockA{\textit{Department of Computer Science} \\
\textit{NED University of Engineering and Technology}\\
Karachi, Pakistan \\
marvi@neduet.edu.pk}

}
\newwatermark[pages=1-8,color=gray!20,scale=3,angle=-45,xpos=0,ypos=0]{IEEE ICCWS 2022}

\maketitle

\begin{abstract}
The tremendous growth in smart devices has
uplifted several security threats. One of the most prominent
threats is malicious software also known as malware. Malware
has the capability of corrupting a device and collapsing an entire
network. Therefore, it’s early detection and mitigation are extremely important to avoid catastrophic effects. In this work, we
came up with a solution for malware detection using state-of-the-art natural language processing (NLP) techniques. Our
main focus is to provide a lightweight yet effective classifier for
malware detection which can be used for heterogeneous devices,
be it a resource constraint device or a resourceful machine. Our
proposed model is tested on the benchmark data set with an
accuracy and log loss score of 99.13\% and 0.04 respectively. 
\end{abstract}


\section{Introduction}
Malware is one of the biggest threats to the smart devices. According to a recent Open Web Application Security Project (OWASP), malicious software is among the top 5 threats to the smart devices \cite{b29}. A smart device holds different types of information i.e., health, traffic, and agriculture data. Some of the information is sensitive such as health data. Therefore, it is important to protect the smart devices from malicious activities. Moreover, a smart device has limited resources such as memory and processing. It is one of the reasons these devices are an easy target for the attackers \cite{b30} as they cannot run bulky antivirus programs. Therefore, there is a need of a light weight malware detection technique which has capability to be integrated within the embedded devices. In addition to this, the time taken for malware detection should be minimal in order to protect the devices before it gets exploited. 
\begin{figure}[htbp]
\centerline{\includegraphics[width=8cm,height=10cm,keepaspectratio]{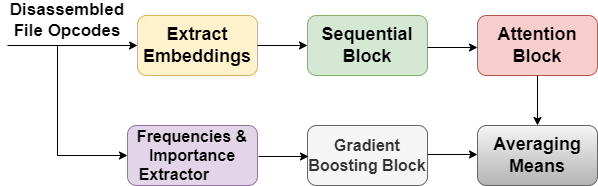}}
\caption{Abstract view of Sequential Embedding-based Attentive Model}
\label{fig21}
\end{figure}
There are three prominent techniques that are used to detect and classify a malware \cite{b18}. Following are the details:
\subsection{Static Analysis (Signature Based)}
In static analysis, program's own structure is compared with the structure of the known malware without considering the execution behavior and other factors. This technique can be helpful where a malware does not have capability to change it's behavior. Nowadays, a malware has a polymorphic and metamorphic characteristics which means that it will get emerge in a powerful manner as it will propagate \cite{b31}. In such cases, it is impossible to detect a malware by static analysis which is purely based on comparing the structural signatures of malware \cite{b25}.
\subsection{Dynamic Analysis (Behavioral Based)}
The dynamic analysis is based on the execution behavior of the program that is running in a sandbox environment. During it's execution, some important metrics are collected such as the system calls a program invokes and file accessing information. These collected metrics are then compared with a malware database and it is declared as malware or benign by using certain threshold levels \cite{b26}. Comparatively, this technique works better than the previous one. Yet, there exists the limitations such as if there is a zero-day malware (due to which most of the systems are compromised) or a new malware family for which the reference metrics are missing, the system would not be able to identify the malware.
\subsection{Heuristic-based}
The above two methods are used for the detection of known malware families. However, the new threats are emerging rapidly and mostly fall into the category of unknown malware families. The heuristic based methods has capability to capture the unknown malicious codes. Recent advances in Machine Learning have created opportunities to use heuristic techniques for the detection and classification of polymorphic as well as metamorphic malware. In this technique, when a binary arrives in a system, it is passed through a Machine Learning (ML) or Deep Learning (DL) pipeline for classification into malware or benign \cite{b27} buckets. This technique has the potential to learn the internal representations and distributions of Malware data and thus enabling the system to detect new malware and prevent zero-day attacks. 

Inspired by the pros of heuristic based techniques, we have used the similar in this research. Here, the machine learning pipeline uses a mix of pre-processing techniques, and machine learning algorithms to predict the probability of the incoming sequence belonging to a malware family as shown in figure \ref{fig21}.

Furthermore,  we use a benchmark data set, "Microsoft BIG 15" which was released by Microsoft on Kaggle\cite{b24}. The dataset comprises of nine different malware families with highly imbalanced classes for the classification problem. With that, following are the major contributions of this research work:
\begin{itemize}
    \item To build up an efficient classifier (both computationally and time-wise) and in doing so we also aim to analyze the impact of class imbalance problems on the malware data.
    \item To compare the results of our proposed method with the existing state of the art methods which uses the same data set.
    \item In addition to this, we also report the results on test data by submitting the proposed model on the Kaggle \cite{b24} platform.
\end{itemize}

\section{Related Work}
In the past years, different Machine learning, deep learning, and Natural language processing based techniques have been used for pre-processing and identifying the presence of a malware in a network of devices. The process includes a thorough analysis of the files sent to the devices and identifying whether the file is malware or benign.

The authors \cite{b19} have used sequential pattern mining technology to detect the most frequent pattern of the opcode sequence for malware identification. In an other similar research \cite{b20}, to identify and categorize a malware, the behavior sequence chain of some malware families has been constructed, and the similarity between the behavior sequence chain and the sequence of the target process has been calculated.

Signature based methods are being used in past for detecting malware in various devices such as mobile \cite{b28}.
The signature-based method is usually involved in detecting already present malware from the past and it includes matching the most similar structure with a known malware but it is insufficient to detect malware that has remained undetected in the past. The process includes first extracting the features from malware of a known family and putting it in a database which is also known as a signature database, so whenever a new program needs to be classified as malware or benign first the features are extracted and then cross-checked and compared from the existing signature database \cite{b21}\cite{b22}. 

Antivirus vendors have been using signature-based detection methodology for years, and it is very quick and effective in detecting known threats. This method is commonly employed to detect malware from the same family. However, it has limitations for detecting newer malware that employs obfuscation and polymorphic methods \cite{b23}.

Malware binary, usually with a file name extension of “.exe” or “.bin”, is a malicious program that could harm computer operating systems. Sometimes, it may have many variations with highly reused basic patterns. This implies that malware binaries could be categorized into multiple families (classes), and each variation inherits the characteristics of its own family. Therefore, it is important to effectively detect malware binary and recognize possible variations. A malware binary file can be visualized to a gray image.

\section{Data Exploration}
In this section, we present the patterns and behaviors of a different malware families using exploratory data analysis techniques.
\subsection{Exploratory Data Analysis}
\subsubsection{Class-wise Data Samples}
\begin{figure}[htbp]
\centerline{\includegraphics[width=8cm,height=8cm,keepaspectratio]{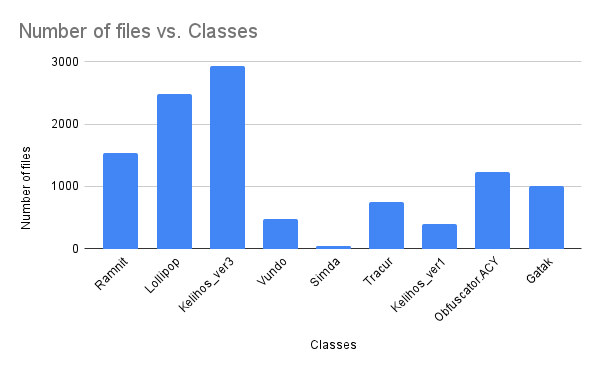}}
\caption{Number of samples per class plot.}
\label{fig1}
\end{figure}

In this part we have performed an analysis on the dataset as a whole. Each family has an opcode sequence and each malware file has an Id as a 20 character hash value that uniquely identifies the file.  Fig. \ref{fig1} represents the number of files amongst different families.

\subsubsection{Opcode-based relation}
\begin{figure}[htbp]
\centerline{\includegraphics[width=8cm,height=8cm,keepaspectratio]{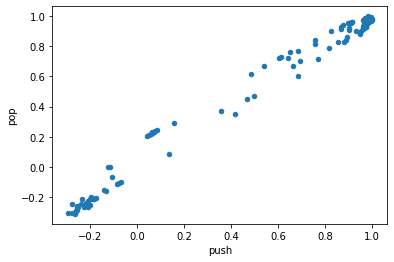}}
\caption{Scatter plot of pop vs push occurrence in the samples.}
\label{fig2}
\end{figure}

\begin{figure}[htbp]
\centerline{\includegraphics[width=8cm,height=8cm,keepaspectratio]{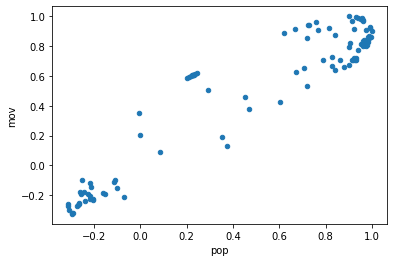}}
\caption{Scatter plot of mov vs pop occurrence in the samples.}
\label{fig3}
\end{figure}

Here, we have calculated the frequency of different opcodes and tried to find a relationship between them. These relations can be linear or non-linear. It is observed that in any class if an opcode \(push\) occurs so \(pop\) also has the same probability of occurring. It is represented using a scatter plot as shown in Fig. \ref{fig2} and Fig. \ref{fig3}.

\subsubsection{Embedding Clusters}
\begin{figure}[htbp]
\centerline{\includegraphics[width=8cm,height=8cm,keepaspectratio]{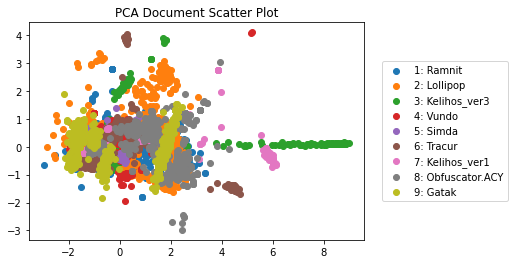}}
\caption{Pricipal Component Analysis components of document embedding.}
\label{fig4}
\end{figure}

\begin{figure}[htbp]
\centerline{\includegraphics[width=8cm,height=8cm,keepaspectratio]{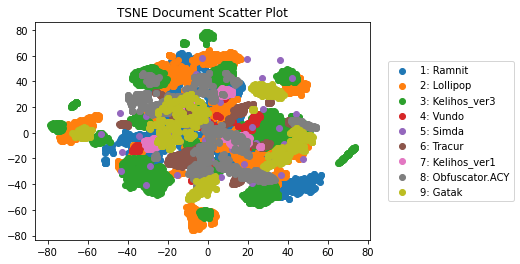}}
\caption{t-SNE components of document embedding.}
\label{fig5}
\end{figure}

In this part, we explore the similarities and differences between clusters of different families that are formed from the learned embedding of the malware opcodes. We train our model on the training corpus, with the opcodes as tokens. Each opcode is treated as a word. Then the document embedding vector is formed using the average of tokens in that document. Fig. \ref{fig4} and Fig. \ref{fig5} shows the plots of Principal Component Analysis (PCA) \cite{b9} components and t-SNE \cite{b8} components of those 100 dimension document embedding.

Although the plots does not show us the totally separable clusters among other classes, One of the possible reasons is that these are the 2-dimensional projections of 100-dimensional vectors therefore, there are high chances  of some information loss while projecting these vectors.

\subsection{Correlation of opcode among each file sample}
\begin{figure}[htbp]
\centerline{\includegraphics[width=8cm,height=8cm,keepaspectratio]{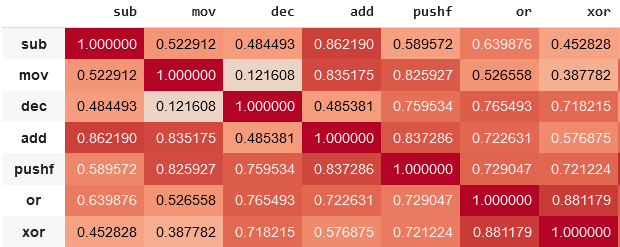}}
\caption{Correlation of opcodes between each other.}
\label{fig6}
\end{figure}

Here, we found the pearson correlation \cite{b10} between different opcodes amongst each other as shown in (Fig. \ref{fig6}). It is quite helpful as it let us know the strength of opcodes' relation with each other. the darker the color the highest the positive correlation.

\section{Methodology}
\begin{figure*}
    \centering
    \includegraphics[width=19cm,height=20cm,keepaspectratio]{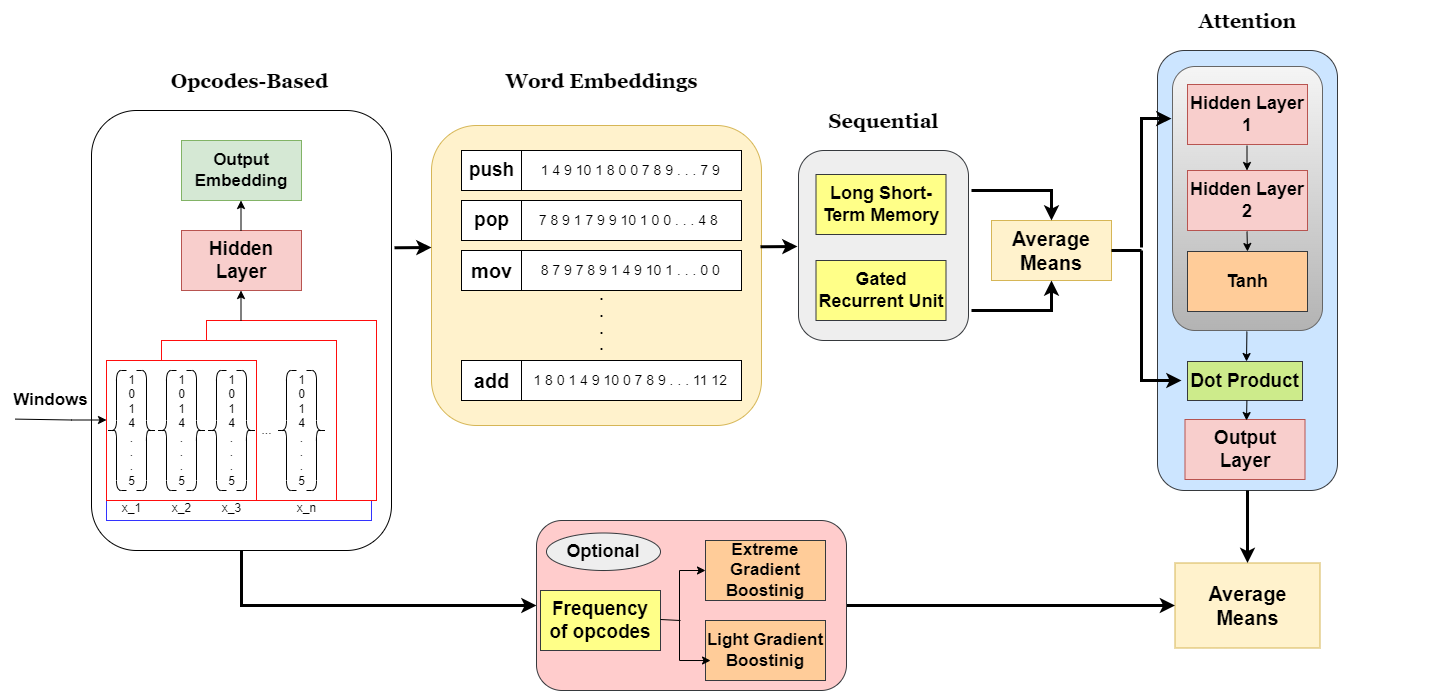}
    \caption{Sequential Embedding-based Attentive Classifier: Our proposed Model}
    \label{fig:arch}
\end{figure*}

Figure \ref{fig:arch} depicts the proposed methodology, the goal is to make an efficient (in terms of time and computing) yet effective classifier for malware classification. Performance in terms of metrics and the time efficiency in real time scenarios are often inverse to each other. Here, for each part, we define a rule of thumb: 1) Use suitable method at each stage. 2) Use an efficient alternate of that method to decrease the time consumption. Secondly, a classifier which can easily identify the hidden patterns without being biased towards the majority class.

\subsection{Window Sliding}
Window sliding propagates from left to right in a sequential manner. However, at a particular step at a time stamp \(t\) it considers only the information contained in a window.\\
For each of the window, the tokens at time stamp \(t\) 
are passed to a network. It learns embedding for each of the contained opcodes and the window passes the sequence covering and learning all the opcodes embedding.

\subsection{Learning Opcodes Embedding}
We focus majorly on learning two things: 1) Context 2) Semantics. Both might be the same in some contexts but here by “Context” we mean to capture the information of opcode i.e., in what particular context it is being used. For example, as we know opcode \(push\) is being used for pushing to stack but the same \(push\), when used with some \(register\), can lead to intentionally generating an error. So by context, we target to capture this contextual information.\\
While the semantics means to know the information or to learn a special representation vector regarding that particular opcode, so based on our previous example, we will be learning a vector
representation that represents the opcode \(push\).\\
The sliding window technique is being used to consider \(n\) continuous sequence at a time \((n=5)\). At time \(t\) we consider the sequence from \(k\) to \(k+n\), where, \(k\) is the starting point. The considered sequence in a particular window is passed through a hidden layer to learn its hidden representations. This process is continued till it reaches the end of the sequence.\\
During testing, same sliding window method is used and the learnt parameters of the hidden layer are used to generate embedding for each opcode in the sequence. The \(d\) dimensional embedding is averaged and complete embedding of whole sequence is also obtained.
\subsection{Sequential Block}
The embedding learned in the previous part is fed into sequential blocks. As we already know, those learned embedding preserve both the contextual and the semantic information which means
that we can learn the information contained in longer sequences. As our data is already massive so there will be many long term dependencies between opcodes, maybe some starting code has some linkage with the conclusive code. In order to capture such dependencies, we have some state of the art architectures already published, one of them is Long Short Term Memory (LSTM) \cite{b1}. Each sequence embedding is passed to multiple LSTM cells.\\
On very basic level, the embedding are passed through the input gate, in order to discard the irrelevant information contained in each of the embedding. We are going to to obtain a probability in range of \(0\) to \(1\) indicating the importance of each scalar value (\(0\) being least important and \(1\) being most important) using this, the irrelevant information is discarded. It is done by applying a \(sigmoid\) function activation on the input embedding, this is having a scalar level importance score. Now, this vector is \(multiplied\) with the output of the last cell to preserve only the important information. Next, we are going to learn the information in longer term dependencies. The obtained vector in the last part having only the useful information can be further used to add new information to it. The new information can be added in  such a way that a single scalar value can either be very important or not important at all, for that we are going to apply \(tanh\) activation function to obtain
a value from range \(-1\) to \(+1\) indicating how much each scalar is important for the sequence. This vector is multiplied with input vector and the scalars contained in it are now representing new information and a new vector \(C_t\) is obtained. This new vector can be added to the useful information preserved vector obtained through the forget gate. Finally the output is passed through an output gate and the learnt vector is first passed to the next cell in the form of \(C_t\) and it is passed as a next state after going through \(tanh\) activation function.\\
Once the embedding are passed through the process, the dimensions will be reduced while preserving and capturing long term dependencies. \\
As we know data is big and sequence contains many information, the model can’t capture all of the information through single LSTM process.In order to add more diversity and capture those dependencies which are ignored by LSTM, we require another architecture in parallel. Here, we can add another LSTM but that would be a compromise on time efficiency as each of the cell have 3 gates, if we reduce the number of gates, we can have an efficient alternative. Luckily, we
have GRU \cite{b2} already published having 2 gates only while performing same as LSTM in our case.
\subsection{Averaging Mean Values}
From the previous part, two different feature vectors collected. The motive of both of them is to represent the information regarding opcodes representation, long-term relations and hidden contextual patterns contained in the sequences.\\
Now in order to get a single feature vector which benefits from the divergence of two feature vectors, we cascade their outputs to the same dimensions, each of the scalar in that dimension is
supposed to represent the same part of the data.\\
For example, if scalar \(i\) in feature vector at position \(j\) produced by GRU represents the data at location \(k\) to \(k+m\). (\(k\) is the starting position and \(m\) is the continuous sequential range) then it is expected that scalar \(h\) produced by LSTM at feature vector’s position \(j\) will also represent the
same information of data at location \(k\) to \(k+m\). \\
Using this method we got the same and diverse representation of the same space. In order to greatly benefit from this divergence, we combine them in such a way that their means converge to the center while eliminating the extreme values. Taking a simple
mean of these will simply center their mean and this will try to eliminate or at least smooth the extreme values.
\subsection{Attention Block}
This is one of the most important part of our model’s architecture where we focus more on what’s important and forget the not-so-important information.\\
Suppose a sequence having 30,000 opcodes and assume that \(98\%\) of the code is normal and the rest of \(2\%\) is suspicious or something which can produce a mal-function in the code. Through our previous block (Sequential Block) we have the patterns information for complete sequence which includes \(98\%\) of good code and \(2\%\) of bad code. Now here is the question: is that \(98\%\) useful for us in differentiating the particular sequence from other
sequences? Obviously that \(2\%\) is more important for us but as that \(2\%\) is in minority it can easily get ignored and our model ends up focusing on the majority part of the code. To tackle this
issue, we have to assign very little attention to \(98\%\) and a high amount of attention to that \(2\%\).\\
To achieve our motive, we are here using a self-attention mechanism which is comparatively very light weight and captures the important details. The reason we did not use the multi-head attention is that it can lead to very expensive computations hence killing the motive
of accurate but efficient classifiers.\\
Taking input from the sequential block, the attention block first prepares an attention vector by passing the embedding through hidden layers and applying \(tanh\) activation function
to it. The reason we applied the \(tanh\) is that it will give outputs in the range \(-1\) and \(+1\) which represents the importance of a particular vector in \(n-dimensional\) space. The dot product is taken between the attention vector and the extracted embedding. It is passed to the output layer to obtain the final outputs.\\
However, the proposed model architecture is finished here but it can also be combined with other models easily using the average means method that we have described above.

\section{Results}
This section includes the details of experimental setup and the results of our proposed approach. Moreover, it also adds the comparison of the proposed approach with the state of the art approaches.
\begin{figure*}
    \centering
    \includegraphics[width=19cm,height=20cm,keepaspectratio]{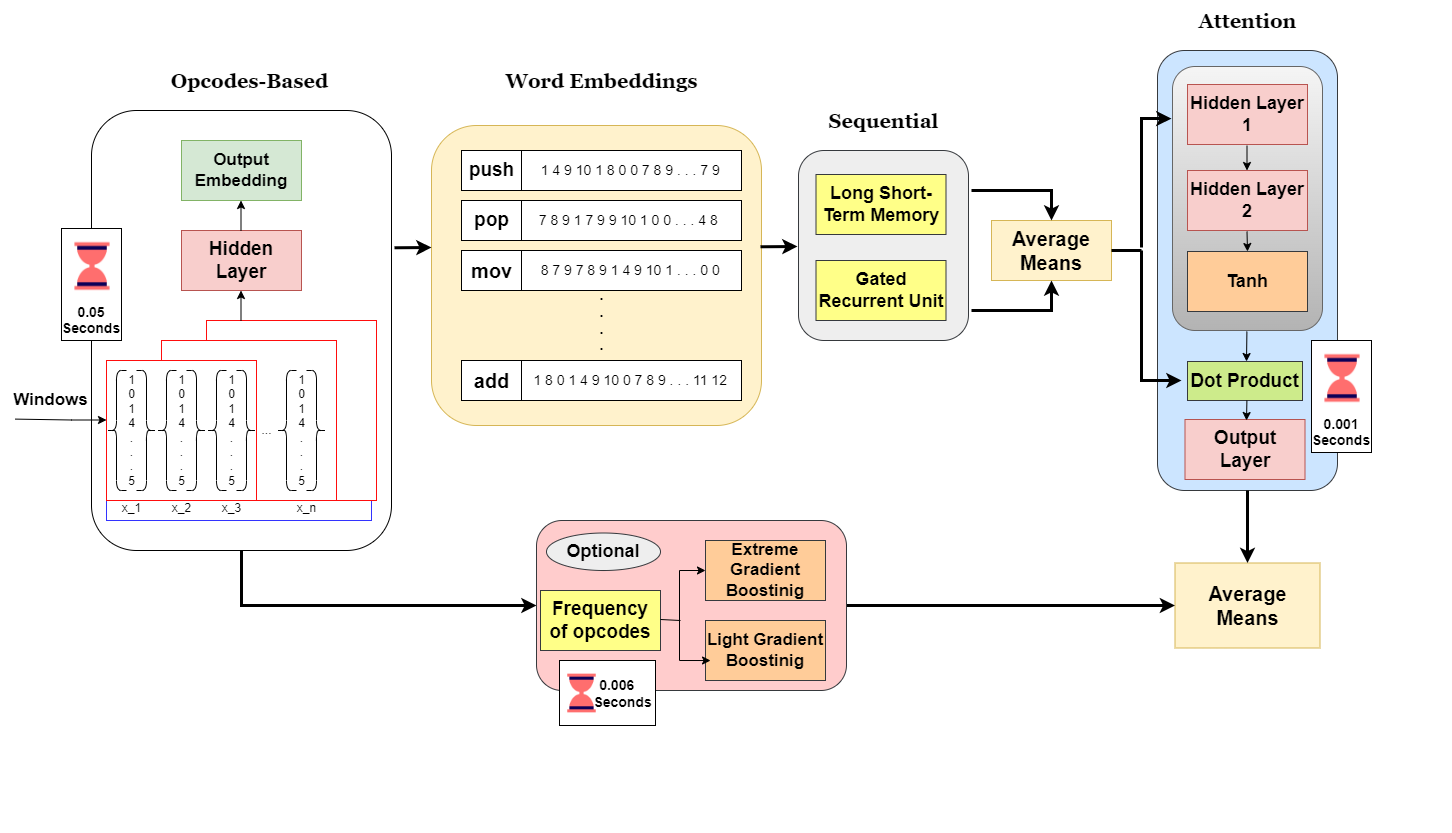}
    \caption{Time taken by each part of the network.}
    \label{fig:time}
\end{figure*}
\subsection{Experimental Setup}
In order to perform the experiments, a pipeline is created as following:
\begin{itemize}
    \item An executable (.exe) file is transferred as an input to the framework.
    \item A reverse engineering process is used to convert the .exe file to an assembly (.asm) file.
    \item The opcodes are extracted from .asm file without disturbing their sequence of opcodes.
    \item At the end, the opcodes are given as an input to the proposed classifier which will construct embeddings and later on use them for classification.
\end{itemize}
In this research, we have used tesla T4 GPU with 16GB RAM to train the model. 
\subsection{Comparative results with other classifiers}
\begin{figure*}
\centerline{\includegraphics[width=20cm,height=18cm,keepaspectratio]{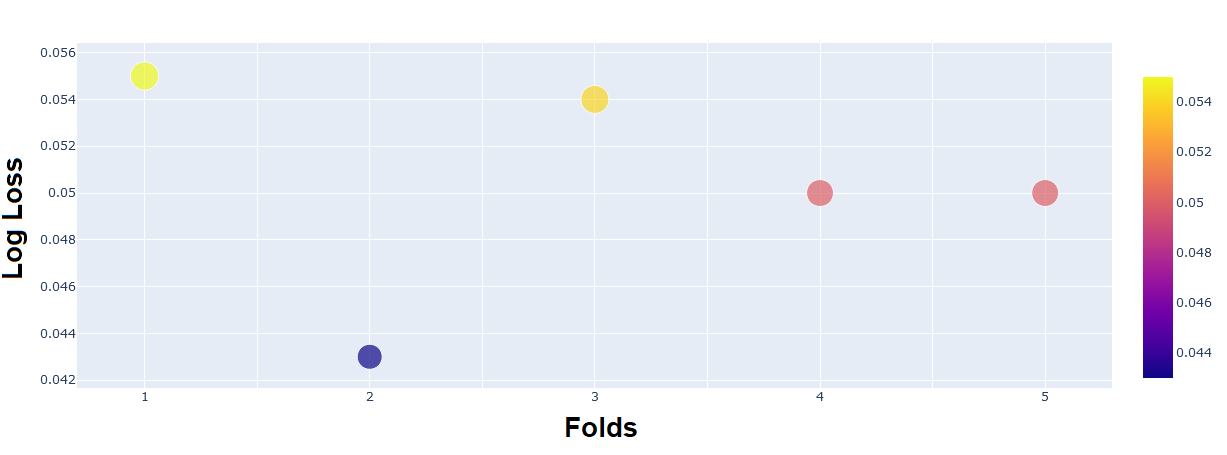}}
\caption{Logloss on 5-fold validation.}
\label{fig9}
\end{figure*}

For validation of the classifier, our primary metric is log-loss on 5-fold validation data and log-loss on Kaggle’s testing data.\\
For the 5 fold validation, the model was trained on 50 epochs on each fold. For all the folds, the loss was between 0.065 and 0.045 (can be seen in Fig \ref{fig9}) which is very similar to the existing methods while being the efficient one.

Other than logloss, we also computed some other metrics such as ROC-AUC score, accuracy and f1 score and compared them with existing methods as shown in \ref{tab1}.

\begin{table}[htbp]
\caption{Comparison of SEA with state-of-the-art models}
\begin{center}
\begin{tabular}{|c|c|c|c|}
\hline
\textbf{Models}&\multicolumn{3}{|c|}{\textbf{Metrics}} \\
\cline{2-4} 
& \textbf{\textit{Accuracy}}& \textbf{\textit{F1 Score}}& \textbf{\textit{Logloss(Kaggle)}} \\

\hline
\textbf{SEA} &0.9912 &\underline{0.9908} &0.0431  \\
(Proposed Method) &  &  & \\
\hline
\hline
Hierarchical CNN \cite{b5}&0.9913 &0.9830 &0.0419  \\
\hline
\hline
CNN with &0.9828 &0.9830 &0.0750  \\
Structural Entropy \cite{b6}& & & \\ 
\hline
\hline
HYDRA \cite{b3}&0.9975 &\textbf{0.9951} &-  \\
\hline
\hline
Orthrus \cite{b4}&0.9924 &0.9872 &-  \\
\hline
\hline
Bytes Files Classifier \cite{b7}&0.9861 &0.9719 &0.3677  \\
\hline
\end{tabular}
\label{tab1}
\end{center}
\end{table}

\begin{figure}[htbp]
\centerline{\includegraphics[width=8cm,height=8cm,keepaspectratio]{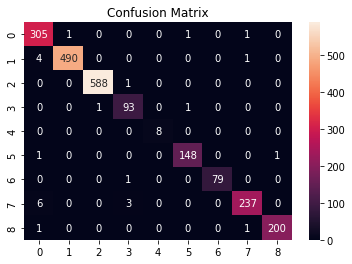}}
\caption{Confusion Matrix based on 5-folds}
\label{fig10}
\end{figure}

Figure \ref{fig10} shows the confusion matrix based upon 5-folds cross validation. It shows that, our model is successful in capturing the patterns of minority classes as well. Even though only a few examples of minority classes were given, still our classifier successfully identified the majority of them. It indicates that our proposed method can give proper attention to hidden patterns among different classes.

\subsection{Time efficiency}
Another metric for our evaluation was the time efficiency of our classifier which makes it the fastest classifier among other methods.We have divided our model’s time into different parts which helps us to understand the time consumption of each component.
\subsubsection{Training Time}
Embedding Training
\begin{itemize}
    \item 100 epochs took 16 minutes 46 secs with 1223 samples on Tesla GPU Neural Network Training
    \item 250 Epochs in 25 minutes on Tesla GPU
    \item 250/25 = 10 Epochs per minute
\end{itemize}
\subsubsection{Time efficiency in real time}
It took our model 686 seconds to process 11K massive data in 5 folds and each of the fold took 137.2 seconds. It means that \textbf{Each sample took 0.0124 seconds.}\\
\begin{itemize}
    \item 0.05 Seconds in embedding extraction
    \item 0.005 seconds in machine learning helping models
    \item 0.001 seconds in neural network part
\end{itemize}
The time taken in real time scenarios that is for test data on each part of the network is shown in Fig \ref{fig:time}.

\begin{table*}[htbp]
\caption{Comparison of trainable parameters of different methods with efficient proposed method.}
\begin{center}
\begin{tabular}{|c|c|}
\hline
\textbf{Models}&{\textbf{Trainable Parameters}} \\

\hline
\textbf{SEA} &\textbf{812,030}  \\
\hline
\hline
Hierarchical CNN \cite{b5}&20,863,302   \\
\hline
\hline
CNN with Structural Entropy \cite{b6} &900,148 \\ 
\hline
\hline
HYDRA \cite{b3}&2,666,617  \\
\hline
\hline
Orthrus \cite{b4}&973,989  \\
\hline
\hline

\end{tabular}
\label{tab2}
\end{center}
\end{table*}

In table \ref{tab2}, we compared trainable parameters of our proposed method with other methods. The reported numbers show that our proposed method has less number of trainable parameters. It makes SEA a suitable choice to be used in light weight smart applications. Although, in terms of logloss, hierarchical CNN was slightly better than our method and in terms of f1-score, HYDRA was slightly better. However, if we compare the number of parameters we can see a huge difference, we were able to almost equal them with very less number of parameters. Moreover, in terms of f1-score and logloss our method has outperformed most of them. Hence, the proposed method is a perfect balance between performance and efficiency.

\section{Conclusion and Future Work}
In this work, we proposed a light weight malware detection technique, called as SEA, for the detection of malicious software in smart applications. By formulating the problem of detection of malicious executable as a highly imbalanced multi-class classification, we developed an attention based model to improve the results on the minority classes as well.
Furthermore, after benchmarking the developed model against state-of-the-art models, we found that our proposed model is comparatively better not only in terms of accuracy and log loss but the number of parameters and time consumption as well. The other factors such as energy consumption, implementation on hardware devices such as Raspberry Pi or ESP32 and delay in the network are the part of future work. 

\section*{Acknowledgment}
This work has been fully supported by Higher Education Commission (HEC) of Pakistan under the Grant No. NRPU-5946

\end{document}